\documentclass[amsmath,amssymb,aps,prl,twocolumn,showpacs,10pt,floatfix,superscriptaddress]{revtex4-2}
\usepackage{graphicx}
\usepackage{dcolumn}
\usepackage{bm}
\usepackage{dsfont}
\usepackage[utf8]{inputenc}
\usepackage[colorlinks,linkcolor=blue,citecolor=blue,urlcolor=blue]{hyperref}
\usepackage[usenames,dvipsnames,svgnames,table]{xcolor}

\usepackage{braket}
\usepackage{bm}
\usepackage{bbold}
\usepackage{amsmath}
\usepackage{amsthm}
\usepackage{mathrsfs}

\begin{document}

\def\equationautorefname~#1\null{Eq.~(#1)\null}
\renewcommand{\figureautorefname}{Fig.}
\def\re{\mathrm{Re}}
\def\im{\mathrm{Im}}
\def\alphare{\alpha_{\re}}
\def\alphaim{\alpha_{\im}}
\def\betare{\beta_{\re}}
\def\betaim{\beta_{\im}}
\title{Dissipation-induced superradiance in matter coupled to a self-interacting cavity}
\newcommand{\jdp}[1]{\textcolor{teal}{#1}}
\newcommand{\seb}[1]{\textcolor{purple}{#1}}
\newcommand{\oz}[1]{\textcolor{orange}{#1}}

\author{Sebastian Schmid}
\affiliation{Institute for Theoretical Physics, ETH Zurich, 8093 Z{\"u}rich, Switzerland}
\affiliation{Department of Physics, University of Strathclyde, Glasgow, United Kingdom}
\affiliation{Department of Physics, University of Konstanz, 78464 Konstanz, Germany}
\author{Matteo Soriente}
\affiliation{Institute for Theoretical Physics, ETH Zurich, 8093 Z{\"u}rich, Switzerland}
\author{Oded Zilberberg}%
\affiliation{Department of Physics, University of Konstanz, 78464 Konstanz, Germany}
\author{Javier del Pino}
\affiliation{Departamento de Física Teórica de la Materia Condensada and Condensed Matter Physics Center (IFIMAC),
Universidad Autónoma de Madrid, E-28049 Madrid, Spain}

\begin{abstract}
Light-matter interactions are often modeled via the Dicke model, namely, by two-level systems coupled to a cavity mode. Alas, the threshold for superradiance is often experimentally inaccessible or hindered by light's diamagnetic term. Here, within the Dicke setting, we consider self-interacting light in a cavity, modeled by a photonic Kerr nonlinearity. We show that negative Kerr nonlinearity gives rise to a low-threshold superradiant phase with spin inversion. While unstable in a closed system, cavity dissipation stabilizes this lit phase, opening avenues for lasing and bath-engineered phases. 
\end{abstract}

\maketitle

Lasing is one of the most prominent examples of light-matter phenomena~\cite{haken1985laser}. It involves many emitters driven by incoherent light, which are placed in a cavity for stimulated emission. Despite the out-of-equilibrium gain channel, i.e., the incoherent drive, the laser produces a coherent output as filtered by the cavity. By constrast, superradiance yields coherence from a collective ground state with a lit cavity, driven by coherent light-matter coupling. Superradiance is effectively described using the Dicke model, where the large ensemble of emitters takes the form of two-level systems that are coupled to an electromagnetic cavity mode~\cite{Dicke1954}. 

In the thermodynamic limit, the Dicke model anticipates a superradiant phase transition (PT)~\cite{hepp1973superradiant,gross1982superradiance,piazza2013bose}, where a large number of emitters become correlated via the cavity, and their emissions constructively interfere. In equilibrium, this transition provides a canonical example of a spontaneous $\mathbb{Z}_2$ symmetry-breaking quantum PT, tuned by a Hamiltonian parameter and driven by quantum fluctuations, making the Dicke model a key framework for collective quantum phenomena and quantum information processing~\cite{Scully2015,Soriente2021}. However, whether such a PT is experimentally achievable has been the topic of much debate and often requires an unrealistically large light-matter coupling strength~\cite{knight1978super,rzaewski1991stability,nataf2010no,Vukics2014,Garbe2017,DeBernardis2018a}. Nonetheless, in a driven-dissipative setting, the superradiant phase prevails and is robust to photon loss, which led to its realization in a plethora of physical systems, including quantum dots~\cite{Scheibner2007, Tighi2016}, cold  atoms~\cite{PhysRevLett.89.253003,PhysRevLett.104.130401,Bien2012, Ritsch2013,brenn2013,Chitra2015,Roof2016,marino2016driven,solano2017,Soriente2018,Soriente2020b,zhu2019dicke,Soriente2021,Ferri2021,mivehvar2021cavity,lin2022dissipation,song2025dissipation,mikheev2025prethermalization}, trapped ions~\cite{aedo2018,lv2018}, and circuit quantum electrodynamics~\cite{mlynek2014,jaako2016}. 

\begin{figure}[t!]
    \centering
    \includegraphics[width=0.8\columnwidth]{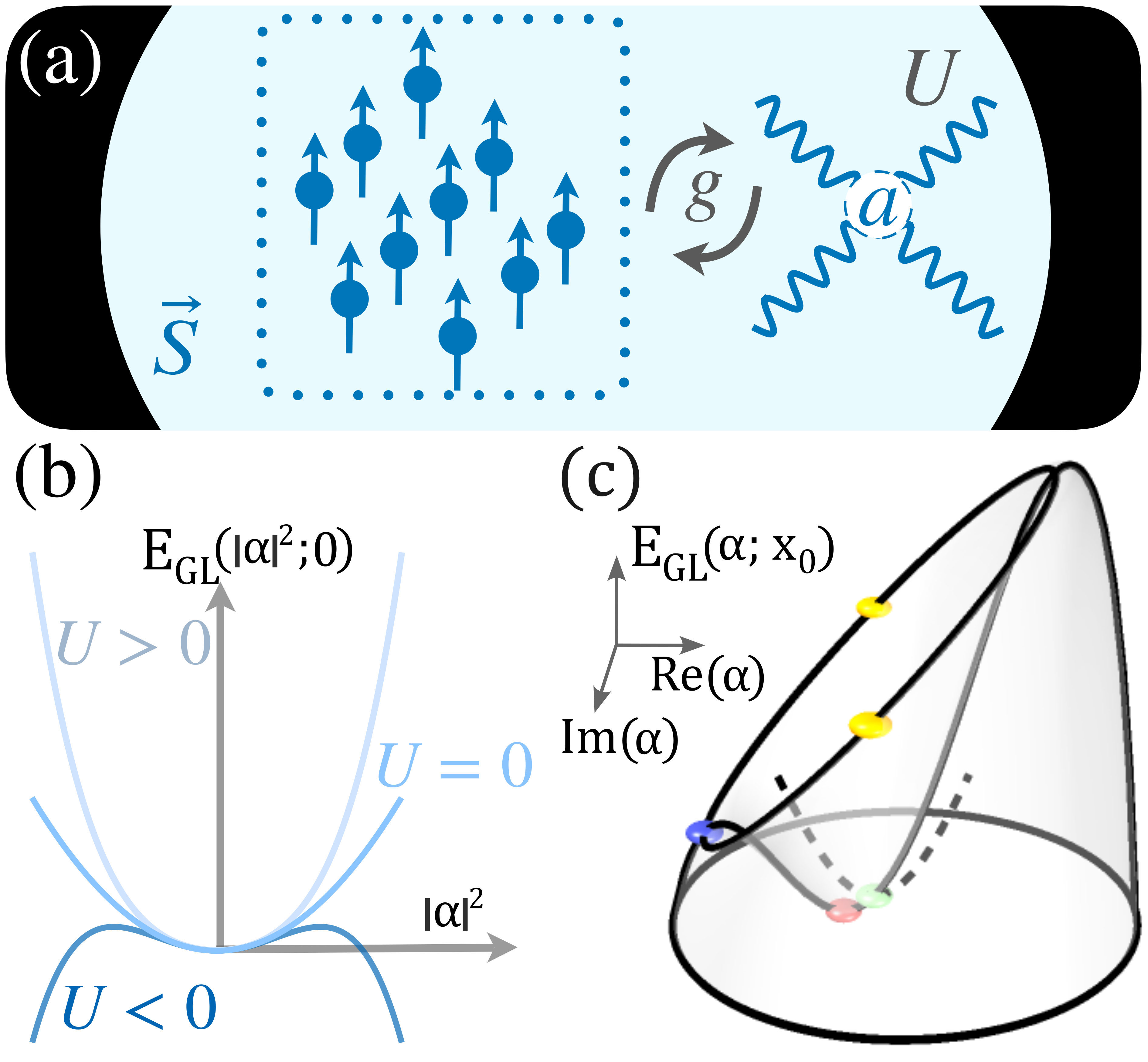}
    \caption{\textit{System.} (a) Kerr-Dicke model [cf.~Eq.~\eqref{eq:Hamiltonian}]: $N$ two-level emitters (blue arrows) couple with strength $g$ to a Kerr cavity with 4-wave mixing strength $U$. (b) Mean-field cavity potential per emitter, $V(\alpha)=\omega_0|\alpha|^2+U|\alpha|^4$, for $U>,=,<0$ [cf.~Eq.~\eqref{eq:mean-field}], whose $|\alpha|$ dependence gives circular $U(1)$ symmetry in the complex $\alpha$ plane. (c) For $U<0$, a spin excitation $x_0$ tilts the potential through $x_0\operatorname{Re}(\alpha)$, selecting an axis and reducing $U(1)$ to $\mathbb{Z}_2$. The dashed curve shows the reversed-$x_0$ partner. The tilt yields a lower saddle (blue dot), two rim points (yellow dots), and, near the bottom, normal (green) and superradiant (red) Dicke phases.}
    \label{fig:setup}
\end{figure}

\begin{figure*}[t!]
    \includegraphics[width=\textwidth]{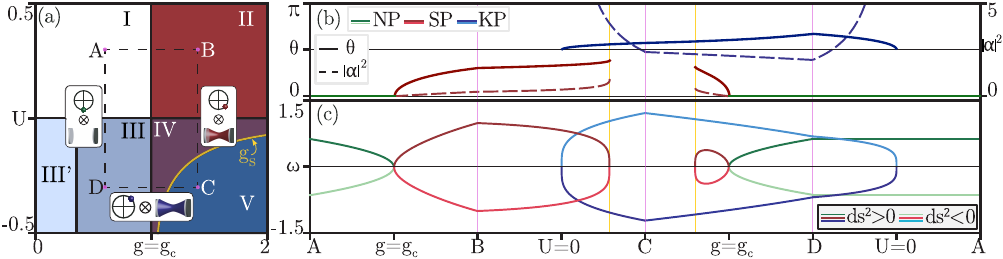}
    \caption{\textit{Closed system phases}. (a) Phase diagram of the closed Kerr-Dicke model~\eqref{eq:Hamiltonian} versus coupling $g$ and Kerr strength $U$, with regions I--V. Insets: state $\xi_k$, showing one $\mathbb{Z}_2$-equivalent spin orientation on the $x$-$z$ great circle and the cavity state (on/off) for NP, SP, and KP. (b) Order parameters $\theta_k$ (solid) and $|\alpha_k|^2$ (dashed), and (c) soft-polariton branch $\omega$ [cf.~Eqs.~\eqref{eq:mean-field}, \eqref{eq:fluc}], along the dashed cut in (a). Colours denote phases. In (c), dark/light hues encode particle/hole character, $\mathrm{ds}^2>0/<0$~\cite{Soriente2020b,dumont2024energy}. Subregion III$^\prime$ lies within region III and is not crossed in (b,c). Other parameters are $\omega_0=\omega_z=1$.}
    \label{fig:closed_pd}
\end{figure*}
Kerr nonlinearities are omnipresent in photonics~\cite{boyd2008nonlinear,Zilberberg2023}. They manifest as an intensity-dependent refractive index of a material~\cite{elsasser2004optical}, which causes self-interaction for light, via four-wave mixing, leading to optical bistability, self-focusing, self-phase modulation, and frequency comb generation~\cite{boyd2008nonlinear}. In cold atom experiments, Kerr effects, sparked by atomic populations, lead to light-matter cooperative manifestation of hysteresis~\cite{martini1993optical,lambrecht1995optical,Gupta2007,gabor2023ground}. Furthermore, effective photonic self-interactions appear in three-level systems in cavities~\cite{phatthamon2024}. At the quantum level, strong photon-photon interactions are crucial to the realization of many-body phenomena in Bose-Hubbard models~\cite{Jaksch1998,Teichmann2009,Dutta2011,hartmann2016quantum,Biondi2017,Collodo2019} and generation of nonclassical states~\cite{PhysRevA.49.R20,PhysRevLett.79.1467,miranowicz1990generation}. Moreover, in superconducting circuits, the intense Kerr effect of microwave radiation enables the realization of qubits~\cite{RevModPhys.93.025005}. Despite their prevalent impact, the role of Kerr nonlinearities in Dicke physics remains mostly unexplored.

In this work, we analyze the effect of an optical Kerr nonlinearity within the Dicke model using a mean-field approach. We evaluate the system both with and without cavity dissipation. Our findings reveal a rich phase diagram that encompasses the standard superradiance transition alongside a novel $\mathbb{Z}_2$ symmetry-breaking phase occurring for negative Kerr nonlinearities. Unlike the standard superradiant phase, this \textit{Kerr-radiant phase} features inverted emitters. It can coexist with the normal phase at low couplings and with the superradiant phase at intermediate interaction strengths. At high couplings, the conventional superradiant phase vanishes completely, and the Kerr-radiant phase remains the sole mean-field solution. Furthermore, while this phase manifests as an unstable excited state in the closed system, it becomes a dissipation-stabilized stationary state once cavity losses are introduced. Crucially, the light-matter coupling threshold for this transition lies below that of the standard Dicke phase. Thus, we predict a route to superradiance that circumvents conventional realization barriers.

 \begin{figure*}[t!]
    \includegraphics[width=\textwidth]{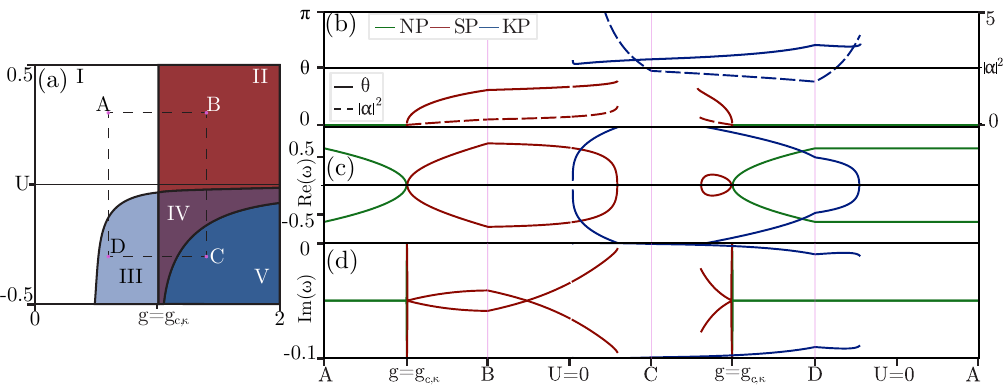}
    \caption{\textit{Open system phases}. (a) Phase diagram of the open Kerr-Dicke model. Regions I and II extend into negative $U$, with the normal phase stable at low $g$. Region III shrinks but remains far below $g_{c,\kappa}$; dissipation removes the imaginary phase and III$^\prime$, while region IV retains SP--KP co-stability. (b) Mean-field order parameters along the ABCD cut. (c,d) Real and imaginary parts of the stability-matrix eigenvalues, giving excitation frequencies and damping. The superradiant PT occurs at $g=g_{c,\kappa}$. Parameters are as in Fig.~\ref{fig:setup}, with $\kappa=0.1$.}
    \label{fig:open_pd}
\end{figure*}

We consider a single-mode Kerr-cavity coupled to $N$ identical emitters [see Fig.~\ref{fig:setup}(a)]
\begin{equation}
    \label{eq:Hamiltonian}
    H = \omega_0^{\phantom{\dagger}} a^{\dagger}a + \omega_z^{\phantom{\dagger}} S_z + \frac{2g}{\sqrt{N}}\left(a + a^{\dagger}\right)S_x + \frac{U}{N} a^{\dagger 2}a^2\,,
\end{equation}
where $a^\dagger$ creates a photon with frequency $\omega_0^{\phantom{\dagger}}$ in the cavity and $\hbar=1$. The $N$ emitters are described as collective spin operators $S_{x,y,z}=(\sum_{j=1}^N \sigma_{x,y,z}^j)/2$, with local Pauli matrices $\sigma_{{x,y,z}}^{j}$ and level spacing $\omega_z^{\phantom{\dagger}}$. The light-matter coupling and cavity's Kerr (quartic) nonlinearity have amplitudes $g$ and $U$, respectively. We assume that the Kerr nonlinearity depends on the light's spatial density similarly to the dependency of light-matter coupling on the matter density. Consequently, we scale the Kerr term by $N$ and keep the model extensive~\cite{hepp1973superradiant}. The sign of $U$ determines if photon interactions in the cavity are repulsive ($U > 0$) or attractive ($U < 0$), see Fig.~\ref{fig:setup}(b).

Our Kerr-Dicke model~\eqref{eq:Hamiltonian} entails a competition between Dicke and Kerr physics.
The standard Dicke model ($U=0, g\neq0$)  hosts a $\mathbb{Z}_2$-symmetry breaking PT from the normal phase (NP, decoupled light-matter with empty cavity) to a superradiant phase (SP, coupled light-matter with lit cavity)~\cite{Dicke1954,hepp1973superradiant,Vukics2014,Chitra2015,Kirton2019,Chiacchio2019}. The $\mathbb{Z}_2$-symmetry arises due to an invariance under the transformation $\{a,S_x\}\to \{-a,-S_x\}$. The symmetry breaks spontaneously when the light-matter coupling exceeds the critical value $g_c = \sqrt{\omega_0^{\phantom{\dagger}} \omega_z^{\phantom{\dagger}}}/2$.
In the opposite limit ($U\neq0, g=0$), we have the Kerr-cavity PTs~\cite{boyd2008nonlinear,Ciuti2016, Biondi2017,Zilberberg2023}. Without light-matter coupling, the cavity has an accidental $U(1)$ symmetry because it depends only on $|\alpha|$. For $U\geq 0$
the landscape has only the empty-cavity minimum $\alpha=0$, while for $U<0$ a circular ridge of unstable extrema. We refer to this unstable, angularly degenerate ridge as an ``anti-Goldstone'' ring [Fig.~\ref{fig:setup}(b)]. The light-matter coupling breaks this $U(1)$ symmetry by tilting the cavity's mean-field potential [Fig.~\ref{fig:setup}(c)]. Relevant for this work: for a negative $U$, the light-matter coupling spontaneously breaks the anti-Goldstone ring (i) into two tilted maxima that coexist with the standard NP and SR  phases (minima) near the empty cavity; and (ii) two ``anti-saddles'' located along the cavity field's imaginary axis which acts as the tilting axis. Note that beyond a critical tilt, the Dicke minima can annihilate with one of the maxima, leaving the system with a single extremal point. 

We begin with analyzing the Ginzburg-Landau (GL) energy potential, $E_{\mathrm{GL}}\equiv \langle H\rangle$, of our model~\eqref{eq:Hamiltonian}. This Gutzwiller mean-field approximation neglects correlations between operators, $\langle AB \rangle \approx \langle A \rangle \langle B \rangle$, while leading to good agreement in the thermodynamic limit~\cite{shirai2013novel,mori2013}. The GL potential depends on the order parameters~\cite{supmat}
\begin{equation}
\label{eq:mean-field}
x = \langle S_x \rangle / N, \quad z = \langle S_z \rangle / N, \quad \alpha = \langle a \rangle/\sqrt{N}\,,
\end{equation}
with macroscopic spin projections $x, z \in [-\frac{1}{2}, \frac{1}{2}]$, and coherent cavity field $\alpha \in \mathbb{C}$. The order parameter $y=\langle S_y \rangle / N$ follows from spin conservation $x^2+y^2+z^2=1/4$. The order parameters span a four-dimensional manifold with coordinates $\xi = \{x,z,\alpha, \alpha^*\}$ corresponding to the $x$-$z$ great circle of a Bloch sphere times the cavity phase space. To obtain the phase diagram of our model, we find the energy's critical points, $\xi_k$, using the saddle-node condition, $\partial E_{\mathrm{GL}}/(\partial \xi)=0$, see Fig.~\ref{fig:closed_pd}(a)~\cite{supmat}. We only consider physical phases, where $\rm{Im}(\xi_k)=0$, and plot their order parameters, see Fig.~\ref{fig:closed_pd}(b).

To classify the critical points, we analyze their Bogoliubov excitations. This is accomplished by (i) expressing the Hamiltonian~\eqref{eq:Hamiltonian} in a coordinate system rotated to align with the spin's elevation angle $\theta_k=\text{arccos}\left(2 z_k\right)$~\cite{supmat}; (ii)  apply the Holstein-Primakoff transformation around the new spin orientation,  $S_{z,k}= \pm \left( b^{\dagger}_k b^{\phantom{\dagger}}_k - N/2 \right)$,  $S_{+,k}=  b^\dagger_k \sqrt{N - b^\dagger_k b^{\phantom{\dagger}}_k}$,
where the bosonic modes $b^\dagger_k$ create excitations away from the $S_{z,k}$ axis~\cite{supmat}; and (iii) expanding $H$ to bilinear order to obtain the excitation Hamiltonian in the thermodynamic limit (large-$N$)~\cite{Baksic2014, Soriente2018, Soriente2020b, Soriente2021}
\begin{align}
\label{eq:fluc}
H_{\rm ex} =& \omega_{c,k} c^\dagger_k c_k  - \Omega_{b,k} b^\dagger_k b_k + U\left[(\alpha_k^*)^2\, c_k^2 + \alpha_k^2 \,(c_k^{\dagger})^2\right] \nonumber\\
&- 2g z_k  \,  (b_k + c_k)\,(b_k^\dagger + c_k^\dagger)\,,
\end{align}
where we defined $\omega_{c,k}= \omega_0 + 4U|\alpha_k|^2 $, $\Omega_{b,k}=2( \omega_z z_k + 4g x_k\, \mathrm{Re}(\alpha_k))$, and $a=\alpha_k\sqrt{N}+c^{\phantom{\dagger}}_k$ with $c^{\phantom{\dagger}}_k$ the bosonic cavity excitation operator. The parameters of the excitation Hamiltonian depend on the specific mean-field solution $k$~\cite{Baksic2014,Soriente2018,supmat}. The excitation spectrum, formed by the complex eigenfrequencies $\omega$, is particle-hole symmetric and pertains to two polariton levels, see Fig.~\ref{fig:closed_pd}(c) for the lower polariton branch. A closing of the excitation gap marks the termination of a given phase. Whether the phase corresponds to a maximum (minimum) of the GL potential is read out of the symplectic norm, $\mathrm{ds}^2$, of the polariton eigenmodes, where $\mathrm{ds}^2> (<)$ 0 indicates particle- (hole-)like excitations, see Fig.~\ref{fig:closed_pd}(c)~\cite{Soriente2021,dumont2024energy}. Hole-like excitations in a closed system are physical, but mark a negative mass instability, i.e., a potential maximum. 

The combined model~\eqref{eq:Hamiltonian} exhibits a rich interplay between Dicke and Kerr physics, see Fig.~\ref{fig:closed_pd}. We find up to twelve critical points, with at most four of them associated with stable phases. Phases with nonzero cavity occupation and spin $x$-component form pairs, differing by a $\pi$ phase shift in the cavity and inversion of the order parameter $x$ due to $\mathbb{Z}_2$ symmetry. When $U>0$, by increasing $g$ along A~$\rightarrow$~B, the system switches from ``region I'' where only the NP exists ($\lvert\theta_{\mathrm{NP}}\rvert = |\alpha_{\mathrm{NP}}| = 0$) to ``region II'' where the SP appears ($\frac{\pi}{2} > \lvert\theta_{\mathrm{SP}}\rvert > 0, |\alpha_{\mathrm{SP}}| > 0$) in a standard second-order Dicke PT. Here, $U$ can only slightly saturate the SP cavity power. Repeating instead the same procedure for $U<0$, at any point in ``region III'', we observe that the NP coexists with a new superradiant phase that is stabilized by the Kerr nonlinearities, i.e., our Kerr-radiant phase (KP). In this exotic phase, the spins are inverted (pointing up), $\lvert\theta_{\mathrm{KP}}\rvert > \frac{\pi}{2}$,  and the cavity is populated $|\alpha_{\mathrm{KP}}| > 0$. Interestingly, the emitters would like to relax by emitting photons into the cavity, while the cavity avoids excessive light absorption by red-detuning away from the matter, cf.~Fig.~\ref{fig:setup}(c). Looking at the symplectic norm of the KP, we observe that it is an unstable state (maximum) in the system. As such, it should experience a negative mass instability akin to an inverted-top, cf.~blue dot in Fig.~\ref{fig:setup}(c). 

For $U<0$, by increasing $g$ along D~$\rightarrow$~C, the NP crosses the Dicke PT at $g_c$ into ``region IV'' where the SP coexists with the KP, whereas at coupling $g=g_S> g_c$ the SP is destabilized, and only the KP remains at ``region V''. We obtain an analytical expression for $g_S$, by identifying that the SP experiences a Kerr-induced instablity due to the cavity detuning away from the matter drive~\cite{supmat}. Echoes of this story appear along different paths in the 2D phase diagram, e.g., at B~$\rightarrow$~C and A~$\rightarrow$~D. Outside the ABCD-cut, at $U<0$ and $g<\omega_z^{3/2}/(4\sqrt{2\omega_0})$, lies ``region III${}^\prime$''~\cite{supmat}. There, in addition to the NP and KP, another $\mathbb{Z}_2$ pair of light-like solutions appear, seemingly uncoupled from matter; they have cavity fields $\alpha_\pm = \pm i \sqrt{-\omega_0 / 2U}$, but $x=y=0$. They arise from the tilting of the cavity potential and are stabilized by the matter fluctuations, see yellow dots in Fig.~\ref{fig:setup}(c). At sufficiently strong light-matter coupling, $g_c>g>\omega_z^{3/2}/(4\sqrt{2\omega_0})$, they destabilize in a transition akin to the NP $\rightarrow$ SR Dicke PT. 

\begin{figure}[t!]
    \centering
    \includegraphics[width=\columnwidth]{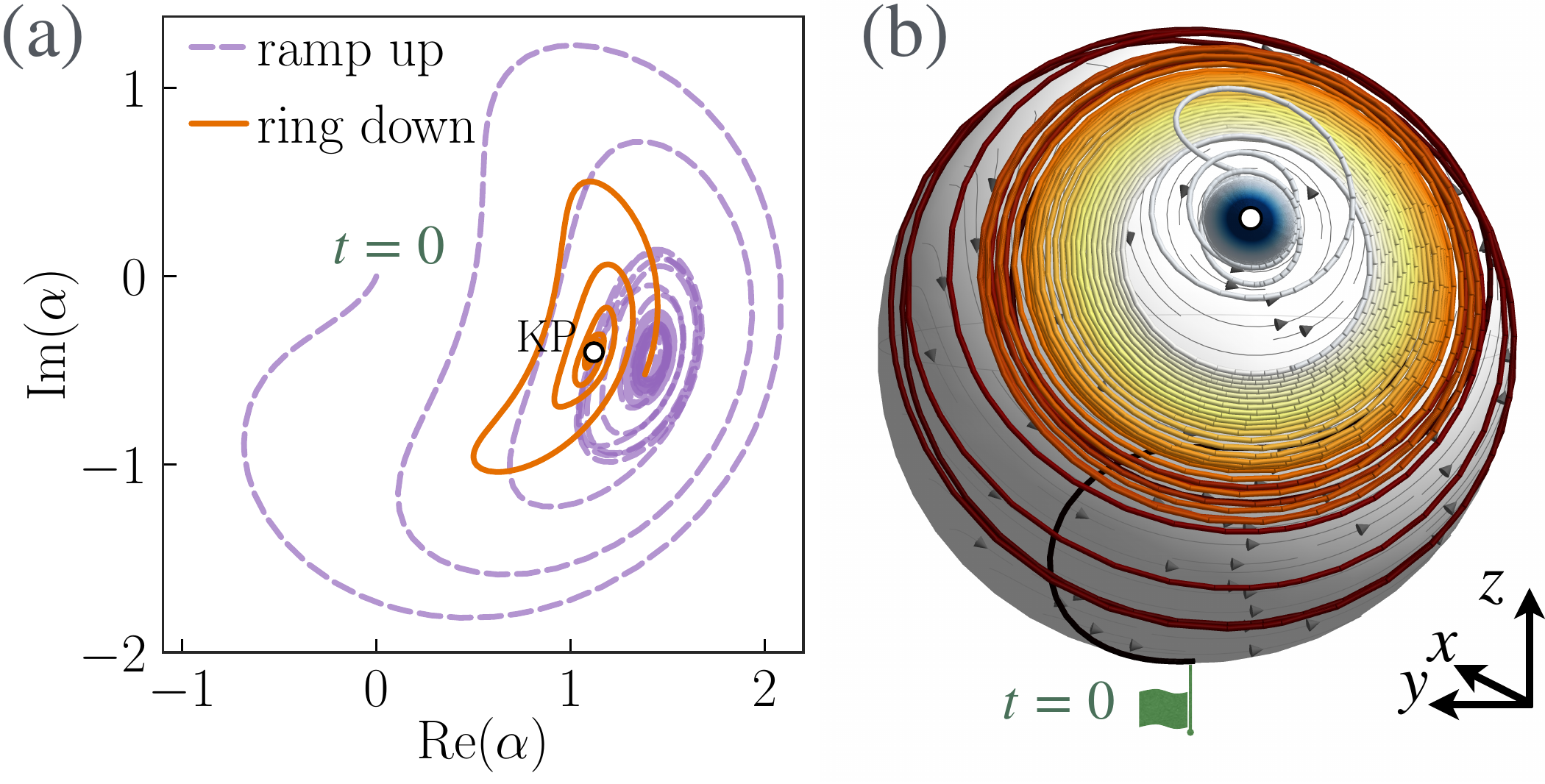}
        \caption{\textit{Preparation of the Kerr-radiant phase (KP)}. The cavity is driven from $\alpha=0$, $z=-1/2$ into the targeted KP using $\zeta=20$, $T_{Pulse}=90$, $g=0.8g_{c,\kappa}$, and $U=-0.45$ [region III in Fig.~\ref{fig:open_pd}(a)]. The ramp (purple) is followed by free relaxation under Eq.~\eqref{eq:operator_eom} (orange). (a) Cavity quadratures. (b) Spin projection on the Bloch sphere, with colors interpolating from purple to orange.}
    \label{fig:4}
\end{figure}

Dicke physics are commonly realized in a driven-dissipative setting, where in a rotating frame relative to the drive, an effective model identical to Eq.~\eqref{eq:Hamiltonian} manifests~\cite{brenn2013,Ritsch2013,Kirton2019,mivehvar2021cavity}. In this setting, photon loss is typically the dominant source of dissipation. To describe the open system, we assume a Markovian bath and introduce losses via the Lindblad master equation for the reduced  density matrix $\rho$ of the system~\cite{breuer2002theory}, with loss rate $\kappa$.
Correspondingly, the dynamics for the operators $A \in \{a, \, S_x, \, S_y, \, S_z\}$ follow Heisenberg's equations of motion
\begin{equation}
    \label{eq:operator_eom}
    \langle \dot{A} \rangle = i\langle[H,A]\rangle + \frac{\kappa}{2}\langle[a^\dagger,A]a+a^\dagger[A,a]\rangle\,.
\end{equation}
We apply the same mean-field approximation as above to Eq.~\eqref{eq:operator_eom}, and solve for the stationary state, $\langle \dot{A} \rangle=0$. Alongside with spin conservation, we obtain the stationary solutions $A_k$. In this case, we analyze stability by separating operators into mean-field and fluctuation parts, $A = \langle A \rangle + \delta A$, solving for real-valued spin-field fluctuations under open-system dynamics. The fluctuation part combines into a stability-matrix with eigenvalues that describe stable states when their $\rm{Im}(\xi_k)\leq 0$. Albeit different from the closed-system Holstein-Primakoff approach, this method smoothly recovers the $\kappa\rightarrow0$ limit~\cite{supmat}.

In Fig.~\ref{fig:open_pd}, we summarize the obtained open-system solutions. Similarly to Fig.~\ref{fig:closed_pd}, we depict in Fig.~\ref{fig:open_pd}(a) and (b) the phase diagram and the order-parameters. In Figs.~\ref{fig:open_pd}(c) and (d), we plot the real and imaginary parts of the (complex) Bogoliubov eigenfrequencies along the path in Fig.~\ref{fig:open_pd}(a), respectively. In the following, we focus on the features that differ from the closed system: for $U>0$, by increasing $g$ along A~$\rightarrow$~B, a superradiant phase transition occurs at $g=g_{c,\kappa}$, at the boundary between regions I and II. The threshold is pushed to $g_{c,\kappa}= \sqrt{\omega_z^{\phantom{\dagger}} \left( \kappa^2 + \omega_0^{\phantom{\dagger}2} \right) / 4\omega_0^{\phantom{\dagger}}}$ due to dissipation~\cite{supmat, brenn2013}. For $U<0$, we observe significant changes from the closed system: (i) Region III shrunk and was taken over by region I, as the KP succumbs to dissipation at sufficient decoupling from the matter's drive; (ii) Nonetheless, at region III (point D), the KP and NP still coexist as in the closed system. The $U$-dependent KP instability boundary between regions I and III obeys an algebraic function, which is crucially appearing at lower values than $g_{c,\kappa}$~\cite{supmat}; (iii) Phase transitions, from region III to IV and from IV to V, occur similarly to the closed system when increasing the coupling along D~$\rightarrow$~C. Note that region IV is slightly eaten away by region II; (iv) The light-like solutions in region III' vanish; since they were decoupled from matter, there is nothing counterbalancing the dissipation and the cavity empties. Remarkably however, the photon loss now stabilizes the negative-mass KP phase, see corresponding $\mathrm{Im}(\omega)<0$ values in Fig.~\ref{fig:open_pd}(c). The stabilization of the exotic KP by dissipation is the main result of this work. Importantly, this stabilization occurs at low light-matter coupling compared with the standard superradiance threshold.

Finally, we discuss the prospects for experimentally observing the newly found phases via: (i) exploit the first-order PT of the KP for $U < 0$: a hysteretic sweep along the C-D direction, while monitoring the spin polarization, reveals a cycle NP~$\rightarrow$~SP~$\rightarrow$~KP~$\rightarrow$~NP. In the KP, spins flip spins from $z=-1/2$ to $z > 0$, leaving a distinct signature; (ii) a time-dependent state-engineering protocol involving initialization of the system at $z = -1/2$ and $\alpha = 0$, and then applying external driving on the cavity defined by 
$\alpha(t) = \alpha_{\mathrm{KP}} \tanh\left( \zeta \frac{t}{T_r} \right)$, with steepness factor $\zeta$,
for $t \in [0, T_r]$, where $\alpha_{\mathrm{KP}}$ is the expected stationary cavity amplitude in the KP. Importantly, this procedure drives the cavity to a stationary amplitude near the KP (i.e., its basin of attraction~\cite{supmat,villa2025topological}), after which ($t>T_r$) the system naturally settles into the KP phase owing to the free evolution in  Eq.~\eqref{eq:operator_eom}, see Fig.~\ref{fig:4}. The cavity evolves on fast timescales set by the ringdown time $\sim \kappa^{-1}$ [cf. Fig.~\ref{fig:4}(a)], while the spin ensemble undergoes sustained precession on the Bloch sphere due to its lack of intrinsic dissipation. This relaxation time mismatch allows an effective description via flows on the Bloch sphere, conditioned on the steady-state cavity amplitude in the KP [cf. Fig.\ref{fig:4}(b)]. Such spin flows are obtained by adiabatic elimination of the cavity mode~\cite{supmat}.

Variations in ramp parameters and initial spin coordinates influence equilibration, see~\cite{supmat}, but two key  features remain: (1) regions in $g-U$ parameter space where the KP reliably initializes, largely irrespective of ramp time, and (2) regions highly sensitive to the final ramp time, where small changes can lead to either the KP or SP. This highlights the system’s extreme sensitivity to the final cavity and spin coordinates after the ramp-up, revealing the intricate structure of its basins of attraction~\cite{villa2025topological}. In the highly sensitive regime, the KP can be stochastically sampled; binning the final ramp times yields a probability distribution, capturing the likelihood of the system ending in the KP or SP phase, providing a statistical characterization of the system as it approaches chaos~\cite{supmat}.

We demonstrate that the Kerr-Dicke phases are experimentally attainable in open quantum systems, thus paving the way to numerous research avenues. Crucially, the presence of negative Kerr-nonlinearity can support dissipation-stabilized superradiant phases and lower light-matter coupling that than of the Dicke threshold. With the correct ramp protocol these phases can be prepared, and a spin population-inverted/lit cavity phase will perpetuate, thus circumventing no-go theorems of superradiance~\cite{knight1978super,rzaewski1991stability,nataf2010no,Vukics2014,Garbe2017,DeBernardis2018a}. 
Beyond the single-photon loss discussed here, dissipative processes are expected to have a significant impact on semiclassical phases. For instance, spin-dephasing can suppress superradiance~\cite{Kirton2017phaseanddecay}, while cavity dissipation tends to maintain it~\cite{bastidas2012noneqqpt, Bezvershenko2021srfluct, Halati2020numerics}. Similar dependency on the specifics of dissipation is found in the Kerr-cavity, where two-photon loss stabilizes different phases to single-photon loss~\cite{Ciuti2016, drummond1980quantum}. To explore quantum effects, future studies can extend to second-order cumulant expansion~\cite{Sanchez2020,Plankensteiner2022} or consider a single emitter inside a Kerr cavity, opposite to the thermodynamic limit~\cite{Werner1991,Gora1992}. Significant effects are also expected after including emitter, e.g. dipole-dipole~\cite{Ostermann2024} or Ising~\cite{Roman2025} interactions, diamagnetic terms~\cite{DeBernardis2018a} and considering multi-level emitters~\cite{Hayn2012}. Investigating non-equilibrium dynamical phases and creating spin squeezing through nonlinearity, as highlighted by the closed-system dynamical matrix of excitations, also offers exciting potential~\cite{Groszkowski2020}. Lastly, studying photon-photon interactions in the Dicke model is crucial for understanding how dynamics cascade into limit cycles~\cite{Keeling2012noneqdyn,Piazza2015,del2024limit} and chaos~\cite{emary2003chaos}.

We thank J.~Marino for critical reading of the manuscript and for fruitful discussions. We acknowledge funding from the Deutsche Forschungsgemeinschaft (DFG) via project numbers 449653034 (Heisenberg), 425217212 (SFB1432), 521530974 (FOR5688), 545605411 (ANR), as well as from the Swiss National Science Foundation (SNSF) through the Sinergia Grant No.~CRSII5\_206008/1.  J.d.P. also acknowledges funding from the Ramón y Cajal program (RYC2023-043827-I), funded by MICIU/AEI (10.13039/501100011033) and FSE+, and from the Spanish Ministry of Science, Innovation and Universities and through the “María de Maeztu” Programme for Units of Excellence in R\&D (CEX2023-001316-M).

\hypersetup{urlcolor=blue}

\hypersetup{urlcolor=blue}

\clearpage

\onecolumngrid
\begin{center}
{\large \textbf{Supplemental Material: Dissipation-induced superradiance in matter coupled to a self-interacting cavity}}
\end{center}
\twocolumngrid

\section{Closed system analysis}
\label{sec:closed_appendix}
In this section, we present detailed calculations for the results on the closed Kerr-Dicke system. We recall that the model describes a bosonic mode, with creation operator $a^{\dagger}$, resonance frequency $\omega_0$, and self-interaction strength $U$, coupled uniformly to $N$ two-level systems. The latter are described by collective spin operators $S_{x,y,z}=\frac{1}{2}\sum_{j=1}^N \sigma_{x,y,z}^j$, with local Pauli matrices $\sigma_{x,y,z}^{j}$ and level spacing $\omega_z$. The Hamiltonian reads ($\hbar=1$)
\begin{equation}
    \label{eq:supp_Hamiltonian}
    H = \omega_0 a^{\dagger}a + \omega_z S_z + \frac{2g}{\sqrt{N}}\left(a + a^{\dagger}\right)S_x + \frac{U}{N} a^{\dagger}{}^2 a^2\,,
\end{equation}
%

We start by bosonizing the spin fields using a Holstein-Primakoff transformation
\begin{equation}
	\label{eq:HolsteinPrimakoff}
	\begin{aligned}
		S_z = \pm \left( b^{\dagger}b - \frac{N}{2} \right), &&
		2S_x = b^{\dagger}\sqrt{N - b^{\dagger}b} + \mathrm{H.c.}\,,
	\end{aligned}
\end{equation}
where $b^{\dagger}$ is a bosonic creation operator. A positive (negative) sign indicates bosonic excitations referenced with respect to the south (north) pole of the Bloch sphere.

\subsection{Mean field phases}
Under the mean-field approximation, we replace operators with coherent states $\sqrt{N} \, \alpha = \langle a \rangle$, $\sqrt{N} \, \beta = \langle b \rangle$, normalized by $N$. This yields the Ginzburg-Landau energy functional
\begin{equation}
E_{\mathrm{GL}}
= \omega_0 |\alpha|^2 + U|\alpha|^4
\pm \omega_z |\beta|^2
+ g(\alpha+\alpha^*)(\beta+\beta^*)\sqrt{1-|\beta|^2}\,.
\end{equation}
Spin conservation entails the factor $\sqrt{1-|\beta|^2}$, limiting the bosonized spin magnitude $\lvert\beta\rvert^2\leq 1$. In polar coordinates, $\alpha = r e^{i\phi}$ and $\beta = s e^{i\theta}$, the energy functional becomes
\begin{equation}
E_{\mathrm{GL}}
= \omega_0 r^2 + U r^4 \pm \omega_z s^2
+ 4grs\cos\phi\cos\theta\sqrt{1-s^2}\,.
\end{equation}

\begin{widetext}
The extrema of this four-dimensional surface in $(r,s,\phi,\theta)$-space are found by setting the partial derivatives to zero:
\begin{align}
\label{eq:r_diff}
0=\frac{\partial E_{\mathrm{GL}}}{\partial r}
&= 2\omega_0 r + 4Ur^3
+ 4g s \cos\phi\cos\theta\sqrt{1-s^2}\,,\\
\label{eq:s_diff}
0=\frac{\partial E_{\mathrm{GL}}}{\partial s}
&= \pm2\omega_z s
+4gr\cos\phi\cos\theta
\left(
\sqrt{1-s^2}
-\frac{s^2}{\sqrt{1-s^2}}
\right)\,,\\
\label{eq:phi_diff}
0=\frac{\partial E_{\mathrm{GL}}}{\partial \phi}
&= -4grs\sin\phi\cos\theta\sqrt{1-s^2}\,,\\
\label{eq:theta_diff}
0=\frac{\partial E_{\mathrm{GL}}}{\partial \theta}
&= -4grs\cos\phi\sin\theta\sqrt{1-s^2}\,.
\end{align}
The normal phase (NP in the main text), defined by $r=s=0$, is always a solution. For $r,s>0$, Eqs.~\eqref{eq:phi_diff} and~\eqref{eq:theta_diff} restrict the angular pairs to two inequivalent families: real solutions with $(\phi,\theta)=(0,0)$ up to the $\mathbb{Z}_2$ and spin-rotation redundancies, and imaginary light-like solutions with $(\phi,\theta)=(\pi/2,\pi/2)$ up to equivalent sign choices.
\end{widetext}

Symmetries further reduce the solution set. A $\pi$ rotation of the $\beta$ field, $\theta\to\theta+\pi$, leaves $S_z$ invariant while flipping $S_x$, and the Dicke $\mathbb{Z}_2$ symmetry maps $(\langle a\rangle,\langle S_x\rangle)\to(-\langle a\rangle,-\langle S_x\rangle)$. Thus, all unique angular pairs reduce to two families: (i) \textbf{Case} $(\phi_1,\theta_1)=(0,0)$, with $\alpha_1,\beta_1\in\mathbb{R}$, leading to
\begin{align}
	\label{eq:r0}
	\frac{\partial E_{\mathrm{GL}}}{\partial r} \Bigg\vert_{\phi=0, \, \theta=0} &= 2 \omega_0 r + 4 U r^3 + 4 g \, s \sqrt{1 - s^2}\,, \\
	\label{eq:s0}
	\frac{\partial E_{\mathrm{GL}}}{\partial s} \Bigg\vert_{\phi=0, \, \theta=0} &\hspace{-1em}= \pm2 \omega_z s
+4gr\left(\sqrt{1-s^2}-\frac{s^2}{\sqrt{1-s^2}}\right)\,, \\
    \intertext{and  (ii)\textbf{ Case} $(\phi_2, \theta_2) = (\frac{\pi}{2}, \frac{\pi}{2})$, where $\alpha_2, \beta_2 \in i\mathbb{R}$ and}
	\label{eq:rpi}
	\frac{\partial E_{\mathrm{GL}}}{\partial r} \Bigg\vert_{\phi=\frac{\pi}{2}, \, \theta=\frac{\pi}{2}} &\hspace{-1em}= 2 \omega_0 r + 4 U r^3 = 2r \left( \omega_0 + 2Ur^2 \right)\,, \\
	\label{eq:spi}
	\frac{\partial E_{\mathrm{GL}}}{\partial s} \Bigg\vert_{\phi=\frac{\pi}{2}, \, \theta=\frac{\pi}{2}} &= \pm2 \omega_z s\,.
\end{align}
With purely imaginary $\alpha_2, \beta_2$, solving Eqs.~\eqref{eq:rpi} and~\eqref{eq:spi} for extrema requires $s = 0$, i.e., no spin-excitation can accompany these solutions. For $U < 0$, such solutions can technically host a nonempty cavity, $r = \sqrt{-\omega_0 / 2U}$ with $z=-1/2$. Yet, cavity and spin are decoupled. This solution corresponds to the light-like imaginary solution (yellow dots in Fig.~1(c) of the main text).Note that the subsequently-derived equations do not hold for this solution, specifically the polynomial relation~\eqref{eq:closed_sol}..

Equations~\eqref{eq:r0} and~\eqref{eq:s0} warrant further simplification.
Consider Eq.~\eqref{eq:r0}: solving it for $s \sqrt{1 - s^2}$ gives
\begin{equation}
	\label{eq:x_sim}
	s \sqrt{1 - s^2} = \frac{-r}{2g} \left( 2Ur^2 + \omega_0 \right)\,.
\end{equation}
The expression $s \sqrt{1 - s^2}$ is reminiscent of the Holstein-Primakoff transformation for $x$.

Consider now Eq.~(\ref{eq:s0}): we set it to zero and multiply by $\sqrt{1 - s^2}$ to obtain
\begin{equation*}
	0 = \pm 2\omega_z s \sqrt{1 - s^2} + 4g \, r \left( 1 - 2s^2 \right)\,.
\end{equation*}
We plug the latter in Eq.~(\ref{eq:x_sim}) and solve for $s^2$, yielding
\begin{equation}
	\label{eq:s_sim}
	s^2 = \frac{1}{2} \left[ 1 \mp \frac{\omega_z}{4g^2}\left( 2U r^2 + \omega_0\right) \right]\,.
\end{equation}
Solving for  $z\equiv \langle S_z\rangle/N$, namely $z = \pm(s^2 - \frac{1}{2})$, we get
\begin{equation}
    \label{eq:z_closed}
	z = -\frac{\omega_z}{8g^2}\left( 2Ur^2 + \omega_0 \right)\,.
\end{equation}

Equation~\eqref{eq:z_closed} shows that a crossover in cavity occupation flips the macroscopic spin to the northern hemisphere:  
\begin{equation}
    \label{eq:z_closed_b}
z \rightarrow
\begin{cases}
\leq 0, & \quad 2U r^2 + \omega_0 \geq 0\,, \\
> 0, & \quad r > \sqrt{\frac{-\omega_0}{2U}}\,.
\end{cases}
\end{equation}
The last condition reflects that spin inversion ($z > 0$) requires $U < 0$.

\subsection{Phase boundary}
\begin{figure}
    \includegraphics[width=\columnwidth]{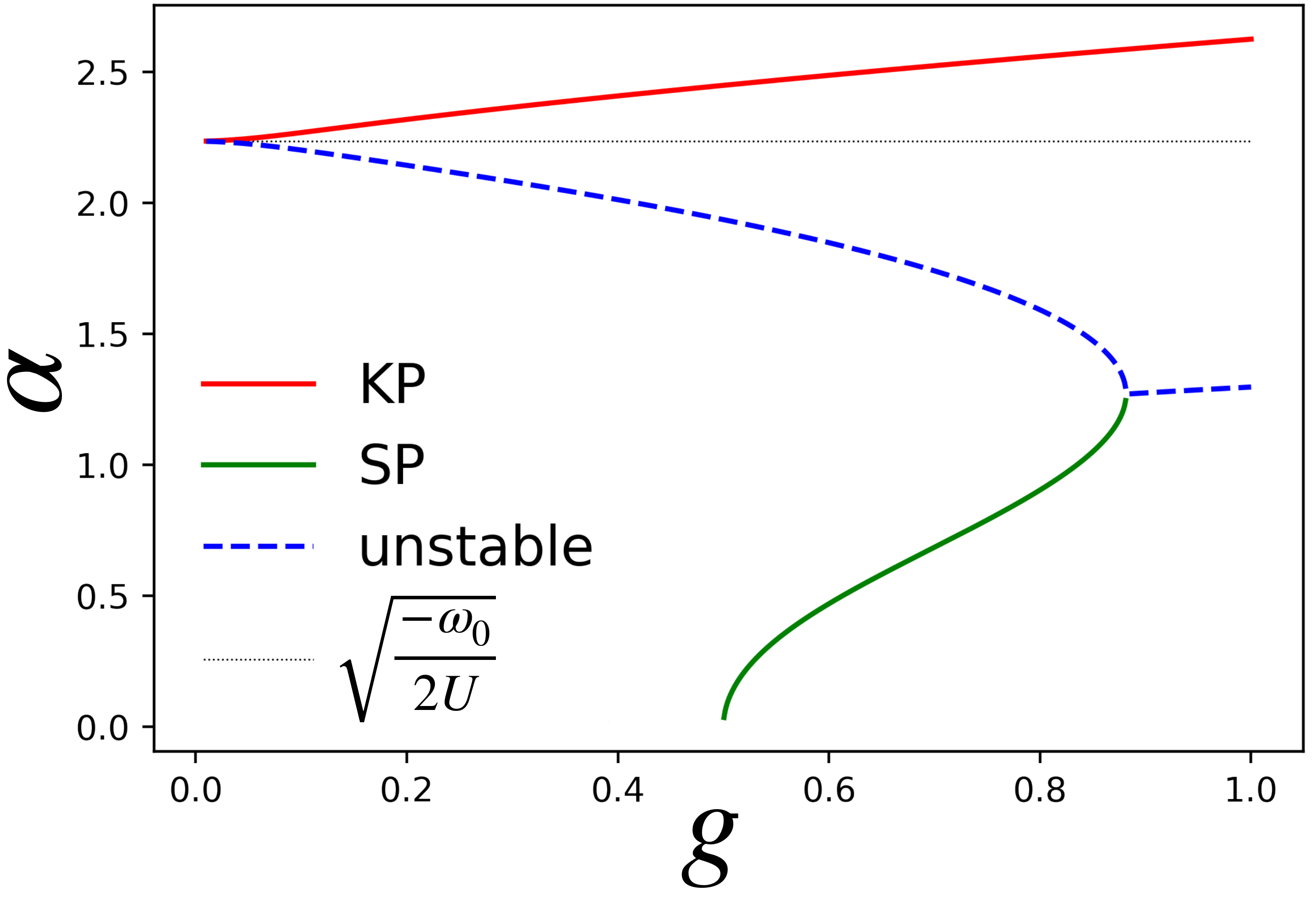}
    \caption{Closed-system cavity occupation $\alpha$ for all physical mean-field solutions at $U=-0.3$. The SP terminates at large $g$ when it merges with an unstable saddle solution. The dashed line, $\alpha=\sqrt{-\omega_0/(2U)}$, marks the occupation at which Eq.~\eqref{eq:z_closed} allows spin inversion. The KP appears only beyond this threshold, consistent with its positive spin polarization.}
    \label{fig:closed_alphas}
\end{figure}

Next to the emergence of a spin inversion, the nonlinearity $U$ has an additional effect. For constant $U < 0$, increasing $g$ will eventually lead the minimum corresponding to the SP to vanish. The phase boundary in $\left( g, U \right)$ parameter space corresponding to the vanishing of the SP can be calculated explicitly. Using~\eqref{eq:x_sim},~\eqref{eq:z_closed}, which enforce spin conservation,  we can derive a 3rd degree equation in $r^2$.
\begin{equation}
    \label{eq:closed_sol}
    \begin{aligned}
        0 = &16 g^2 U^2 \, r^6 + \left( 16g^2U \omega_0 + U^2 \omega_z^2 \right) \, r^4 \\
        &+ \left( 4g^2 \omega_0^2 + U \omega_0 \omega_z^2 \right) \, r^2 + \frac{\omega_z^2 \omega_0^2}{4} - 4g^4\,.
    \end{aligned}
\end{equation}
The solutions $r_k$ to Eq.~\eqref{eq:closed_sol} determine the extrema of the Ginzburg-Landau surface and thus our mean-field closed system solutions. The $\mathbb{Z}_2$ symmetry in solutions is manifest in Eq.~\eqref{eq:closed_sol} containing a polynomial in $r^2$, while the order of Eq.~\eqref{eq:closed_sol} indicates that the NP, SP and KP exhaust all possible mean-field solutions (up to 3 real solutions). Analyzing these solutions reveals that at a critical $g_s$ the SP merges with a saddle-point and they become two complex-valued solutions to Eq.~\eqref{eq:closed_sol}. This mechanism is shown explicitly in Fig.~\ref{fig:closed_alphas}, where the SP branch terminates at a low cavity occupation, while the KP branch, associated with spin inversion, persists at larger occupation. The number of purely real roots is determined by the discriminant $\Delta$. Thus, between $g_c = \sqrt{\omega_0 \omega_z}/2$ and $g_s$, the discriminant of Eq.~\eqref{eq:closed_sol} is positive, while above $g_s$ it is negative. Consequently, at $g_s$, the discriminant is zero. We can calculate the implicit relation $U_s(g)$ analytically as follows. The discriminant is a cubic polynomial in $U$,
\begin{equation}
    \label{eq:discr}
    \Delta (U) = U^3 - 24 g^2 \, U^2 + \left( 1 - 36 g^4 \right) 192 g^4 \, U - 512 g^6\,,
\end{equation}
where in~\eqref{eq:discr} we set $\omega_0 = \omega_z = 1$ for simplicity. Its own discriminant is positive for $g > g_c$, ensuring three real roots. This guarantees the SP phase exists for some $U$ when $g > g_c$. Transforming \eqref{eq:discr} into a depressed cubic, the roots take trigonometric form, with the middle root defining the critical $U_s(g)$ where the SP vanishes,
\begin{equation}
    U_s(g) = 2 \sqrt{\frac{-p}{3}} \cos\left[ \frac{1}{3} \arccos\left( \frac{3q}{2p} \sqrt{\frac{-3}{p}} \right) - \frac{2\pi}{3} \right]\,.
\end{equation}
Here, $p = -27 \cdot \left(2g\right)^8$ and $q = -27 \cdot \left(2g\right)^{10}$.

\subsection{Stability analysis}
The possible phases of the system correspond to the extrema of the energy functional, defined in terms of the coordinates $\xi = \{x, z, \alpha, \alpha^*\}$, where $x$ and $z$ represent the macroscopic spin projections, and $\alpha$ describes the coherent cavity field. We analyze their excitation spectrum by considering fluctuations around each extremum, $\xi_k = \{x_k, z_k, \alpha_k, \alpha_k^*\}$.

To express the fluctuations as a quadratic form, the square root in the Holstein-Primakoff transformation~\eqref{eq:HolsteinPrimakoff} must be expanded into linear and quadratic terms. This expansion includes non-negligible contributions because the spin values in $\xi_k$ can lie relatively far from the north and south poles of the Bloch sphere, which are the anchor points of the Holstein-Primakoff expansion. To address this, we rotate the macroscopic spin in the $xz$-plane, aligning it so that in the new coordinates $x' = 0$ and $z' = \pm \frac{1}{2}$. This transformation ensures that $\beta' = 0$, enabling an expansion around small fluctuations.

For each extremum $\xi_k$, the fluctuations are defined as $b' = \sqrt{N} b_k$ and $a = \langle a \rangle + \delta a = \sqrt{N}(\alpha_k + c_k)$. Retaining terms up to second order in the fluctuation operators results in the quadratic fluctuation Hamiltonian:
\begin{equation}
\label{eq:appendix_fluc_H_rot}
	\begin{aligned}
		H_{\rm fl} =& \left( \omega_0 + 4U|\alpha_k|^2 \right)\, c_k^\dagger c_k + U \left(\alpha_k^*\right)^2\, c_k c_k + U \alpha_k^2 \,c_k^\dagger c_k^\dagger \\
		&+ \left( \omega_z \cos(\theta_k) - 2g \sin(\theta_k)\left( \alpha_k + \alpha_k^* \right) \right) \, b_k^\dagger b_k \\
		&+ g \cos(\theta_k) \left( c_k b_k + c_k^\dagger b_k + c_k b_k^\dagger + c_k^\dagger b_k^\dagger \right)\,,
	\end{aligned}
\end{equation}
with $\theta_k = \arccos\left(-2z_k \right)$. Developing the trigonometric factors, we rewrite Eq.~(\ref{eq:appendix_fluc_H_rot}) as
\begin{equation}
\label{eq:appendix_fluc_H}
	\begin{aligned}
		H_{\rm fl} =& \left( \omega_0 + 4U|\alpha_k|^2 \right)\, c_k^\dagger c_k + U (\alpha_k^*)^2\, c_k c_k + U \alpha_k^2 \,c_k^\dagger c_k^\dagger \\
		&- 2\left( \omega_z z_k + 2g x_k \left( \alpha_k + \alpha_k^* \right) \right) \, b_k^\dagger b_k \\
		&- 2g z_k \left( c_k b_k + c_k^\dagger b_k + c_k b_k^\dagger + c_k^\dagger b_k^\dagger \right)\\
        =& \boldsymbol{\epsilon}_k^{\dagger} \bar{M}_{\rm fl} \boldsymbol{\epsilon}_k\,.
	\end{aligned}
\end{equation}
Here, $\bar{M}_{\rm fl}$ is a coefficient matrix (the $k$ index is dropped) and $\boldsymbol{\epsilon}_k$ is the Nambu spinor, $\boldsymbol{\epsilon}_k^T = \left( c_k^{\dagger} \quad b_k^{\dagger} \quad c_k \quad b_k \right)$. The emerging Bogoliubov normal modes in the system follow from diagonalizing the dynamical matrix $\bar{D}_{\rm fl}$\cite{SM:fetter2012quantum},
\begin{equation}
    \bar{D}_{\rm fl} = \bar{M}_{\rm fl} I_-\,,
\end{equation}
where $I_- = \mathbb{1} \otimes (-\mathbb{1})$.
Using the following abbreviations $\omega_{c,k}= \omega_0 + 4U|\alpha_k|^2 $, $\Omega_{b,k}=2( \omega_z z_k + 4g x_k\, \mathrm{Re}(\alpha_k))$,
the dynamical (Bogoliubov de Gennes) matrix reads
\begin{equation}
	\bar{D}_{\rm fl} =
	\begin{pmatrix}
		\omega_{c,k} & -2g z_k & 2U \alpha^2 & -2g z_k \\
		-2 g z_k & \Omega_{b,k} & -2g z_k & 0 \\
		-2U (\alpha^*)^2 & 2g z_k & -\omega_{c,k} & 2g z_k \\
		2g z_k & 0 & 2g z_k & -\Omega_{b,k}
	\end{pmatrix}\,.
\end{equation}

The eigenvalues of the dynamical matrix describe the response to fluctuations around the mean-field solutions.
The excitations' eigenvectors further provide us with the symplectic norm $\mathrm{ds}_{\mathbf{v}}^2 \equiv \mathbf{v}^\dagger I_- \mathbf{v}$, associated with each excitation eigenfrequency, where $\mathbf{v}$ is an eigenvector of $\bar{D}_{\rm fl}$. The symplectic norm determines the nature of the excitations: it can be either positive ($\mathrm{ds}^2 > 0$) or negative ($\mathrm{ds}^2 < 0$) and is a measure of the particle- or hole-like nature of the excitation, respectively~\cite{SM:Soriente2020b,SM:dumont2024energy}.

\section{Open System treatment}
\label{sec:open_appendix}

We study the open Kerr-Dicke model at the mean-field level, starting from the closed-system case (Appendix \ref{sec:closed_appendix}) and incorporating cavity dissipation. 
We achieve this via the master equation for the density matrix $\rho$,
\begin{equation}
    \label{eq:quantum_master}
     \dot{\rho} = -i \left[ H, \rho \right] + 2\kappa\mathcal{L}_a\left[\rho\right],
\end{equation}
where $H$ is the Kerr-Dicke Hamiltonian~\eqref{eq:supp_Hamiltonian} and
$\mathcal{L}_a[\rho]=a\rho a^{\dagger}-\frac{1}{2}\{a^{\dagger}a,\rho\}$
is the Lindblad superoperator accounting for cavity loss. With this convention, the field amplitude decays at rate $\kappa$.
The resulting equations for the averages of cavity and spin operators read
\begin{align}
    \langle \dot{a} \rangle &= -\left(i\omega_0 + \kappa\right)\langle a \rangle - \frac{2ig}{\sqrt{N}} \langle S_x\rangle - \frac{2iU}{N}\langle a^\dagger a a \rangle\,, \\
    \langle \dot{S_x} \rangle &= -\omega_z \langle S_y \rangle\,, \\
    \langle \dot{S_y} \rangle &= \omega_z \langle S_x \rangle - \frac{2g}{\sqrt{N}}\left(\langle a^\dagger S_z \rangle + \langle a S_z \rangle \right)\,, \\
    \langle \dot{S_z} \rangle &= \frac{2g}{\sqrt{N}}\left(\langle a^\dagger S_y \rangle + \langle a S_y \rangle \right)\,.
    \label{eq:open_eom}
\end{align}
Using the mean-field approximation, i.e., setting $\langle AB\rangle\approx\langle A\rangle\langle B\rangle$ and writing $\alpha =  \langle a \rangle /\sqrt{N}$ for the cavity and $x,y,z = \langle S_{x,y,z} \rangle / N$ for the spin, we look for stationary states ($\langle \dot{Q} \rangle = 0$) and arrive at a set of three equations with only single-operator expectation values.
\begin{align}
	(i \omega_0 + \kappa) \alpha &= - 2ig x - 2i U\alpha|\alpha|^2\,,\\
	y &= 0\,, \label{eq:alpha_st_x}\\
	x  &= \frac{2g}{\omega_z} \left( \alpha^* + \alpha \right) z\,.
    \label{eq:z_and_x}
\end{align}
Together with spin conservation, we collect a closed set of four equations. The solutions to this set are better understood by changing variables to the unnormalized quadratures $X_\alpha = \text{Re}(\alpha)$ and $Y_\alpha = \text{Im}(\alpha)$ 
\begin{align}
    \label{eq:mf_x}
	x &= -\frac{1}{2g} \left[ 2U X_\alpha \left( X_\alpha^2 + Y_\alpha^2 \right) + \omega_0 X_\alpha + \kappa Y_\alpha \right]\,,\\
    z &= \frac{-\kappa \omega_z}{8g^2 X_\alpha Y_\alpha} \left( X_\alpha^2 + Y_\alpha^2 \right)\,,
	\label{eq:mf_z}
\end{align}
with 
\begin{equation}\label{eq:X_alpha}
    X_\alpha = \frac{\kappa \pm \sqrt{\kappa^2 - 8 U Y_\alpha^2\left( \omega_0 + 2UY_\alpha^2 \right)}}{4 U Y_\alpha}\,.
\end{equation}
The last equation for $Y_\alpha$ is found using Eqs.~\eqref{eq:mf_x} and~\eqref{eq:mf_z} together with spin conservation. Note that, since $X_\alpha \in \mathbb{R}$ in Eq.~\eqref{eq:X_alpha}, the discriminant inside the square-root must be positive. This restricts the permissible values for $Y_{\alpha} = Y_{\alpha}(U)$ to a compact interval. Moreover, the negative branch in Eq.~\eqref{eq:X_alpha} reduces in the limit $U \rightarrow 0$, using L'Hôpital's rule, to $X_\alpha = \omega_0Y_\alpha/\kappa$,
while the positive branch has no finite $U\to0$ limit, consistent with its association with the Kerr-induced branch. 
Additionally, taking the limits $\kappa, \mathrm{Im}(\alpha) \rightarrow 0$ in Eq.~\eqref{eq:mf_x} and using Eq.~\eqref{eq:z_and_x}, we correctly recover the relation found in the closed system Holstein-Primakoff treatment, cf.~Eqs.~\eqref{eq:x_sim} and~\eqref{eq:z_closed}.

\begin{figure}
    \includegraphics[width=\columnwidth]{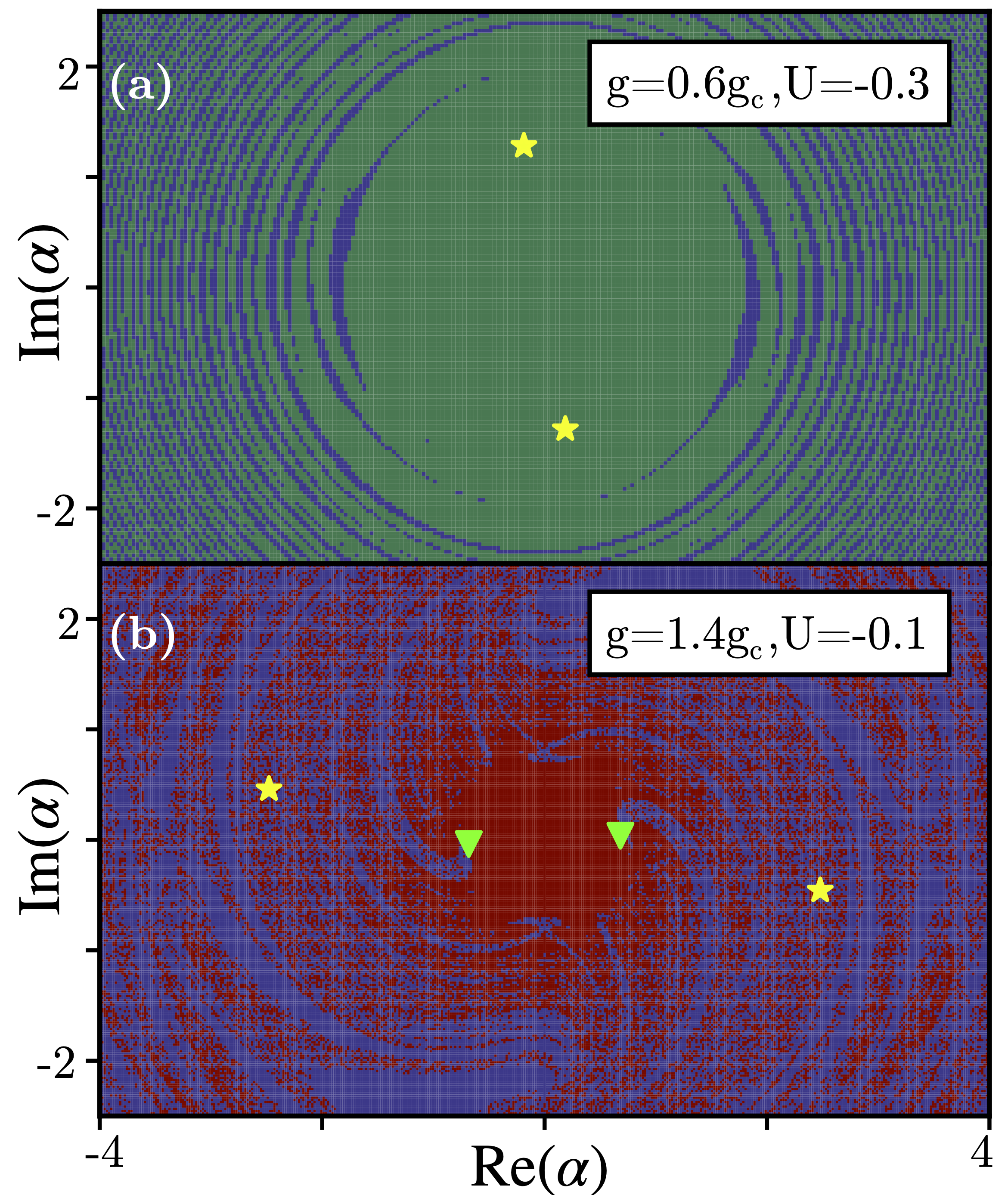}
    \caption{Basins of attraction in cavity phase space of the open Kerr-Dicke model. Basins are color-coded: green for NP, red for SP, and blue for KP. The initial order parameters are $x=y=0$, $z=-\frac{1}{2}$, indicating spins point downwards. In $g-U$ parameter-space (see Fig.~3(a) of main text). (a) is at NP-KP and (b) at SP-KP bistability. Mean-field solutions for cavity-fields are depicted as triangles (SP) and stars (KP). With the finite-size KP basin, the cavity can be excited at the right phase to allow the system to equilibrate towards the KP.}
    \label{fig:basin}
\end{figure}

\subsection{Signatures of the two branches}\label{sec: signature}
From Eq.~\eqref{eq:X_alpha} we can deduce the relation $\text{sgn}(z) = \text{sgn}(X_\alpha Y_\alpha) = - \text{sgn}(U)$ for the positive branch (cf.~Eq.~\eqref{eq:X_alpha} with the $+$ sign before the square root), which for $U < 0$ directly hints at a steady state with spin inversion; this is the Kerr-radiant phase (KP) we have reported on in the main text. Additionally, we then know that any phase of the positive branch type for $U < 0$ has to hold that $X_{\alpha}$ and $Y_{\alpha}$ are  of the same sign. In contrast to this, the negative branch (which is associated with the superradiant solution) has $X_{\alpha}$ and $Y_{\alpha}$ with opposite signs, as long as $U < 0$ and $Y_{\alpha}^2 < \frac{-\omega_0}{2U}$. The latter diverges as $U \rightarrow 0$ from below and is never reached for the superradiant phase.
This means that there is an additional signature distinguishing the SP and the KP. The most prominent one of course is the spin inversion, but also the cavity fields live in different quadrants in the complex plane. $\mathbb{Z}_2$-symmetry implies two solutions for both the SP and the KP, connected via a parity-transformation ${\alpha, x} \rightarrow {-\alpha, -x}$. Thus, in the cavity field, two quadrants host the SP phases, while the other two host the KP phases.

\subsection{Stability analysis}

After calculating the stationary mean-field solutions from Eqs.~\eqref{eq:z_and_x}, we analyze their linear stability. For the open system, we work directly with the real-valued mean-field variables $\boldsymbol{\xi}=(X_\alpha,Y_\alpha,x,y,z)^T$, rather than introducing a Holstein-Primakoff expansion around each stationary spin configuration. This spin-field formulation keeps the open-system equations in the variables used in the main text and avoids the additional axis rotations required in the closed-system Bogoliubov analysis.

To make this explicit, we linearize the mean-field equations around a stationary solution $\boldsymbol{\xi}_k=(X_{\alpha,k},Y_{\alpha,k},x_k,y_k,z_k)^T$ as $\delta\dot{\boldsymbol{\xi}} = J_k \delta\boldsymbol{\xi}$, with
\begin{equation}
J_k=
\begin{pmatrix}
A_k & B_k & 0 & 0 & 0\\
C_k & D_k & -2g & 0 & 0\\
0 & 0 & 0 & -\omega_z & 0\\
-4gz_k & 0 & \omega_z & 0 & -4gX_{\alpha,k}\\
0 & 0 & 0 & 4gX_{\alpha,k} & 0
\end{pmatrix},
\end{equation}
where
\begin{align}
A_k&=-\kappa+4UX_{\alpha,k}Y_{\alpha,k}, \\
B_k&=\omega_0+2UX_{\alpha,k}^2+6UY_{\alpha,k}^2,\\
C_k&=-\omega_0-6UX_{\alpha,k}^2-2UY_{\alpha,k}^2, \\
D_k&=-\kappa-4UX_{\alpha,k}Y_{\alpha,k}.
\end{align}
The eigenvalues $\lambda_n$ of $J_k$ determine the stability of the corresponding phase: a stationary solution is dynamically stable when all physical perturbations decay, namely when the relevant eigenvalues satisfy $\mathrm{Re}\,\lambda_n<0$. The spin-length constraint restricts physical spin fluctuations to the tangent plane of the Bloch sphere, so the radial spin direction is not an independent instability channel. At the mean-field phase boundaries, one or more eigenvalues approach the imaginary axis, producing critical slowing down, or equivalently diverging relaxation times. In the parameter regimes considered here, the loss of stability occurs through soft modes, consistent with the second-order mean-field transitions reported in the main text.

\subsection{Basins of attraction and preparation protocol}

We focus on regions III and IV of Fig.~3 in the main text, where more than one stable stationary state is present. In these bistable regimes, the relevant dynamical question is not only which stationary states exist, but also which one is reached from experimentally accessible initial conditions. The figures in this section provide consistency checks for the preparation protocol used in the main text. Their main message is simple: the Kerr-radiant phase can be reached from extended regions of initial conditions, although the outcome can become sensitive near basin boundaries.

We consider two initialization protocols. First, we prepare the cavity at a chosen initial field ($\alpha_0$), while the spins initially point down, ($x=y=0$), ($z=-1/2$), and then let the system evolve freely. Second, we use the finite-time cavity ramp discussed in the main text, drive the system close to the KP mean-field amplitude, switch off the drive, and follow the subsequent free relaxation. The final state is identified from the long-time cavity occupation and spin polarization, which distinguish the NP, SP, and KP solutions.

\paragraph*{Cavity quench}
First, we analyze quenches from an initially prepared cavity field with deexcited spins ($z=-1/2$). Figure~\ref{fig:basin} shows how attractors in the cavity phase space appear for specific $(g, U)$-values. In Fig.~\ref{fig:basin}, we map out the initial cavity field values $\alpha_0$, with $z=-1/2$, leading to a given equilibrated state after free time evolution. The basins, color-coded by their corresponding stationary phase, show an intricate pattern, where regions leading to the KP phase are adjacent with those leading to NP or SP phases, indicating that small variations in initial cavity fields $\alpha_0$ can result in drastically different outcomes. The cavity mean-field values of the different states (SP and KP) are denoted by markers (triangles and stars). Extended regions of initial conditions leading to the KP exist in both regions III and IV, showing that exact preparation at the stationary point is not required. Near the boundaries between basins, small changes in the initial cavity field can select different final states, as expected in a nonlinear bistable system. The figure therefore supports the ramp strategy used in the main text: by preparing the cavity with the appropriate phase and amplitude, the system can relax into the KP.

\begin{figure*}
    \includegraphics[width=0.7\linewidth]{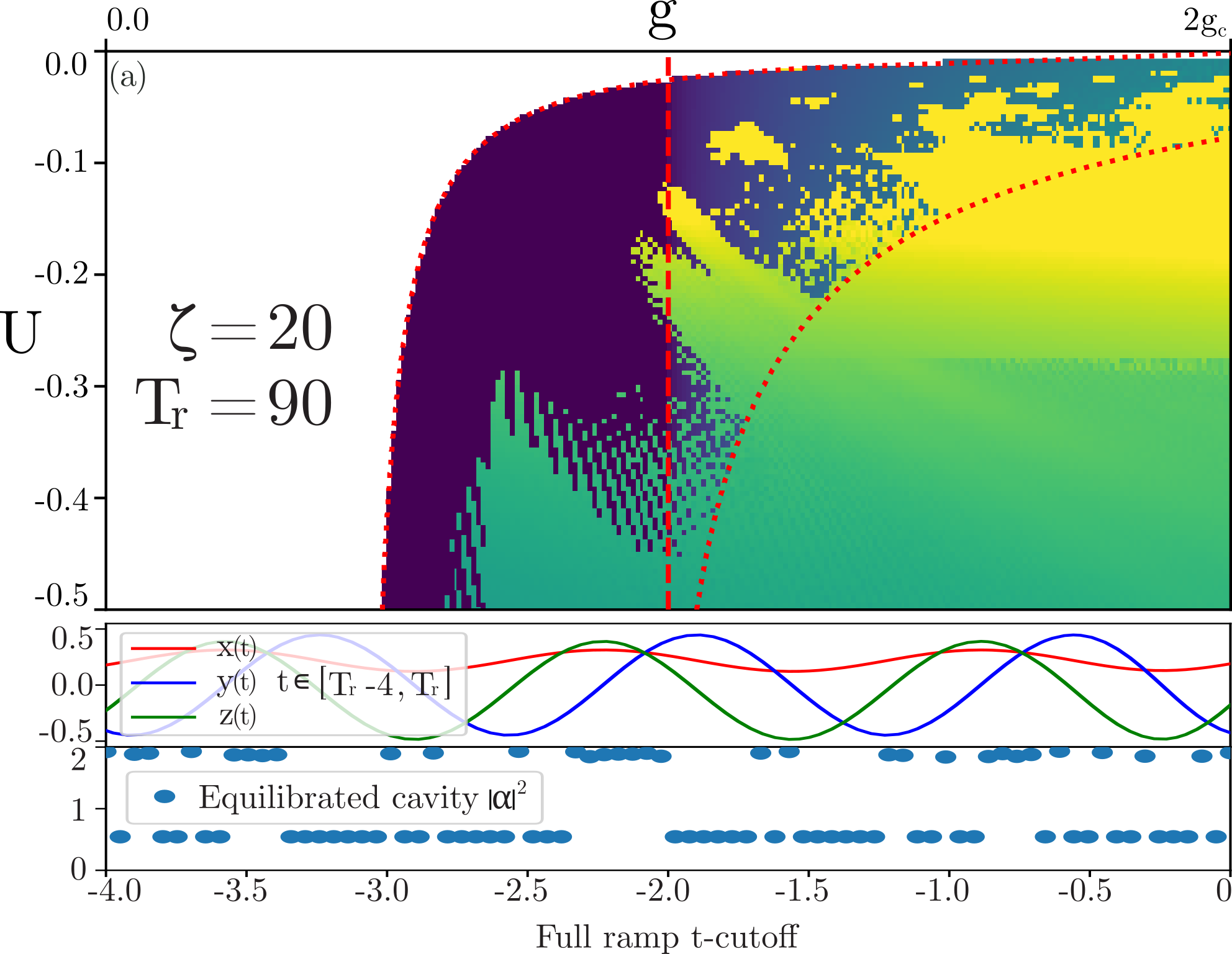}
    \caption{\textbf{Robustness of the ramp protocol in $g-U$ parameter space.}(a) Final cavity occupation $|\alpha(t_f)|$ after a ramp with $\zeta=20$, $T_r=90$, followed by free evolution until $t_f=900$. Large final amplitudes identify trajectories that reach the KP, while smaller amplitudes correspond to relaxation into the NP or SP. (b) Same analysis when the ramp is interrupted earlier, illustrating parameter regions where the outcome is sensitive to the precise ramp duration.}
    \label{fig:ug_basin}
\end{figure*}

\begin{widetext}

\paragraph*{Cavity ramp at different parameters}

We next test the finite-time ramp protocol across the $g-U$ plane. The cavity is driven close to the KP mean-field value according to
\begin{equation}
\alpha(t)=\alpha_{\rm KP}\tanh(\zeta t/T_r),
\qquad 0\leq t\leq T_r,
\end{equation}
and subsequently evolves freely for $t>T_r$. We use the final cavity occupation $|\alpha(t_f)|$ as a practical indicator of the reached phase, since the NP, SP, and KP have well-separated cavity amplitudes.

Figure~\ref{fig:ug_basin} shows that broad regions in both III and IV relax to the KP after the ramp, indicating that the protocol is robust over a finite parameter range. Other regions are more sensitive to the precise ramp duration. This occurs because the cavity can relax faster than the spin, so small changes in the end of the ramp can place the spin on different sides of a basin boundary. The resulting sensitivity is illustrated in Fig.~\ref{fig:rampvar}. Importantly, this does not affect the main conclusion: there are extended parameter regimes where the KP is reliably prepared by a simple cavity ramp.

In Fig.~\ref{fig:rampvar}(a) we bin the angles in the $yz$-plane and show the ratio between equilibrations into the KP and the SP. The resulting peaks identify spin-angle windows from which relaxation into the KP is more likely. Fig.~\ref{fig:rampvar}(b) color-codes spin positions at the end of ramps, indicating the resulting equilibrium state. While the cavity mostly equilibrates, slight oscillations remain due to incomplete spin relaxation, echoing the cavity sensitivity seen in Fig.~\ref{fig:basin}. Most importantly, the stable regimes in regions III and IV identified in Fig.~\ref{fig:ug_basin} are insensitive to these variations, and the KP is faithfully prepared over a wide range of parameters.

\begin{figure}[t]
    \includegraphics[width=\columnwidth]{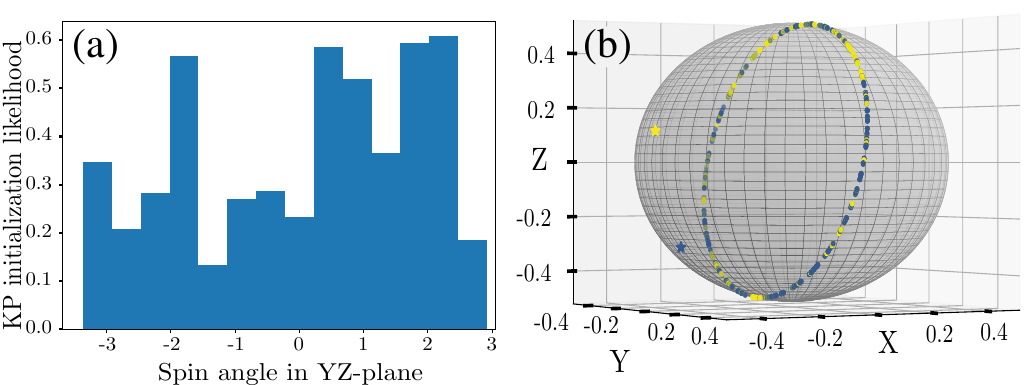}
    \caption{\textbf{Sensitivity to the ramp duration near basin boundaries.} Small changes in $T_r$ can change the spin position after the preparation stage and thereby select different attractors. (a) Ratio of trajectories relaxing to the KP after binning the final spin angle in the $yz$-plane. (b) Spin positions at the end of the ramp, color-coded by the final relaxed state: KP (yellow) or SP (blue).}
    \label{fig:rampvar}
\end{figure}

\end{widetext}

\subsection{Conditional dynamics on the Bloch Sphere}
\begin{figure}[t]
    \includegraphics[width=0.8\columnwidth]{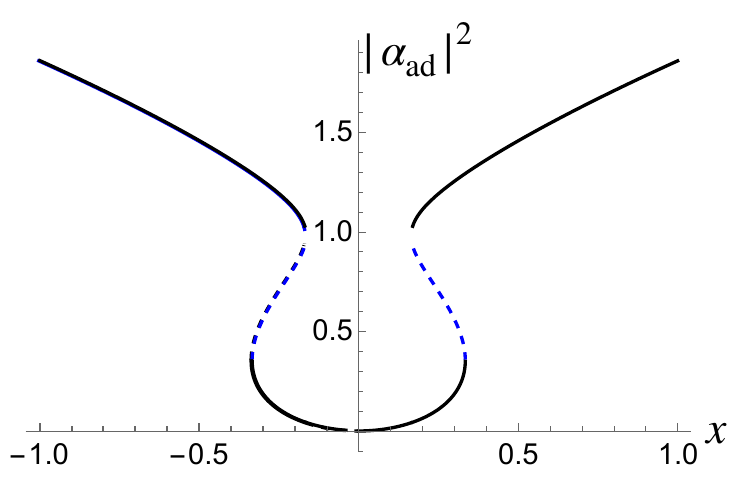}
    \includegraphics[width=0.8\columnwidth]{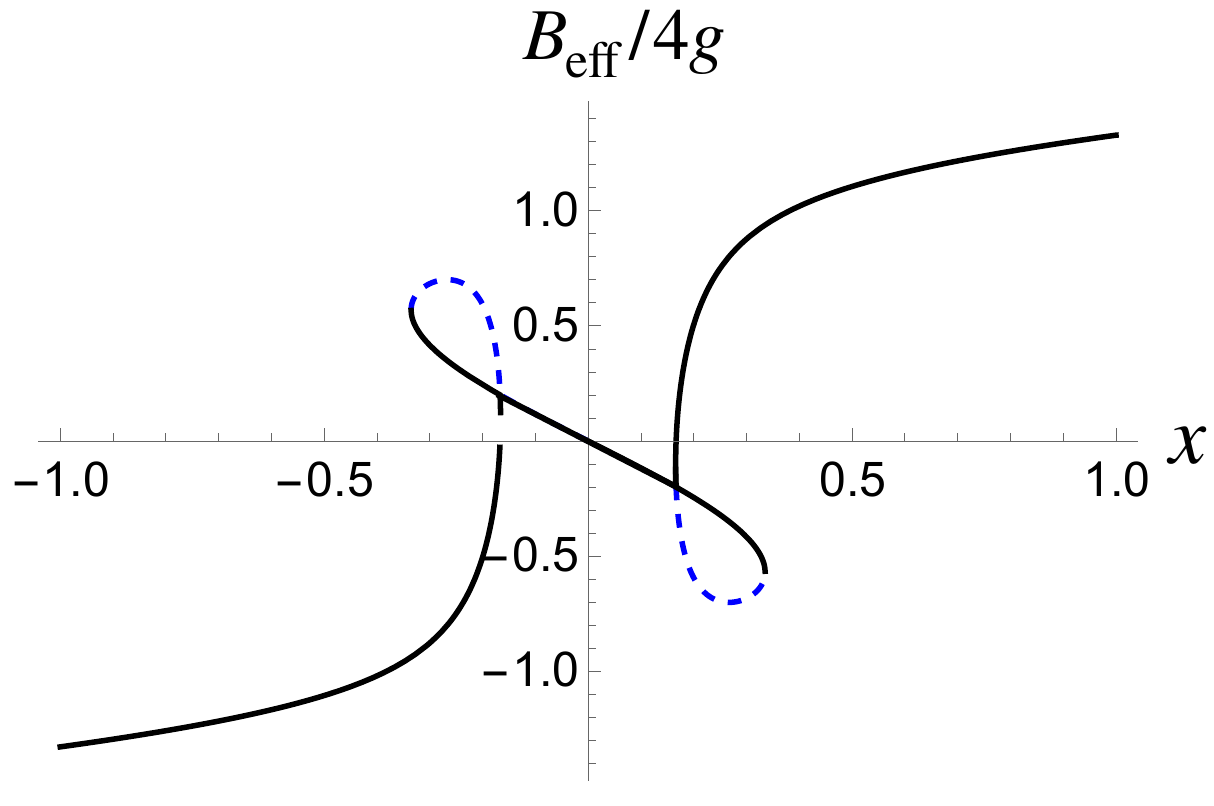}
    \caption{(top) Adiabatic cavity population as a function of the spin order parameter $ x $ for $ \omega_0 = 1 $, $ g = 0.6 $, $ U = -0.5 $, and $ \kappa = 0.2 $. The curve exhibits the characteristic S-shape of a Duffing resonator, reflected under $ x \to -x $, corresponding to a phase shift of $ \pi $ in the effective $ x $-drive. Here solid (dashed) lines correspond to stable (unstable) solutions. (bottom) Effective magnetic field experienced by the spins once the cavity stabilizes.}
    \label{fig:alpha_ad}
\end{figure}

In this section, we describe the mathematical details of the spin-projected system dynamics. The starting point is the set of mean-field equations~\eqref{eq:open_eom}. Since no decay is assumed for the spins, the cavity field evolves on a much faster timescale, $\sim \kappa^{-1}$, allowing for its adiabatic elimination. This means that the cavity field remains ``frozen'' at its steady state, satisfying \eqref{eq:alpha_st_x}. Under this approximation, the cavity behaves as a Kerr resonator driven by the spin component $x$, which can exhibit up to three solutions, namely the real roots of the cubic Eq.~\eqref{eq:alpha_st_x} that have unwieldy analytical expressions, omitted here. Two of such solutions, dubbed $\alpha_{\mathrm{ad}}^{l}(x)$ ($l=1,2$) are dynamically stable, see e.g.,  
Fig.~\ref{fig:alpha_ad}.

The spin order parameters  $x, y, z$  then follow sets of nonlinear differential equations, where  $\alpha$  is replaced by each $\alpha_{\mathrm{ad}}^l(x)$, i.e.:
\begin{align}
    \dot{x} &= -\omega_z y, \\
    \dot{y} &= \omega_z x - 4g\mathrm{Re}(\alpha_{\mathrm{ad}}^l)z, \\
    \dot{z} &= 4g\mathrm{Re}(\alpha_{\mathrm{ad}}^l(x))y.
    \label{eq:open_eom_proj}
\end{align}
Equations~\eqref{eq:open_eom_proj} correspond to Bloch-like dynamics, where the spin precesses around effective, inhomogeneous magnetic fields $\vec{B}^l_\mathrm{eff}(x) = \big( 0, 4g \mathrm{Re}(\alpha_{\mathrm{ad}}^l(x)), \omega_z \big)
$. The field's amplitude and sign depend on the solution: for $x > 0$, the high-amplitude Kerr solution gives $(B_\mathrm{eff}^l)_y(x) > 0$ and the low-amplitude Kerr solution gives $(B_\mathrm{eff}^l)_y(x) < 0$. This pattern reverses for $x < 0$.

\subsection{Cavity Phase-Space Representation}

The mean-field phases can also be analyzed in cavity phase space by eliminating the spin components. Enforcing spin conservation and solving for the stationary spin order parameters, we obtain two branches:
\begin{equation}
x_\pm = \pm\frac{2 g X_\alpha}{\sqrt{16 g^2 X_\alpha^2+\omega_z^2}}, \quad
z_\pm = \pm\frac{\omega_z}{2 \sqrt{16 g^2 X_\alpha^2+\omega_z^2}},
\end{equation}
while $y_\pm =0$. These expressions define the cavity phase space under the assumption that the spins have settled into their stationary amplitudes.
While they do not capture the full system dynamics, they offer an alternative perspective on its behavior. In this approach, the cavity quadratures $X_\alpha\equiv \mathrm{Re}(\alpha)$ and $Y_\alpha\equiv \mathrm{Im}(\alpha)$ obey
\begin{align}\label{eq:cavity_flow_nonlinear}
\dot{X}_\alpha &= (\omega_0 + 2 U |\alpha|^2)Y_\alpha - \kappa X_\alpha  , \\
\dot{Y}_\alpha &= - \left(\omega_0 \pm \frac{4 g^2}{\sqrt{16 g^2 X_\alpha^2 + \omega_z^2}} + 2U|\alpha|^2\right)X_\alpha  - \kappa Y_\alpha.
\end{align}
with $|\alpha|^2 = X_\alpha^2 + Y_\alpha^2$. 

These equations provide a qualitative understanding of the system’s dynamical phases. Equation \eqref{eq:cavity_flow_nonlinear} shows that the cavity undergoes second-order self-coupling (backaction) via light-matter interaction, introducing a nonlinear shift in the $Y_\alpha$ rotation frequency that saturates for large $X_\alpha$, i.e., when the only quadrature coupled to the spins is large. Unlike parametric driving, which modifies both $X_\alpha$ and $Y_\alpha$ \cite{SM:Zilberberg2023}, this backaction, when $U = 0$ and $\kappa = 0$, leads to an SP where the cavity field is linearly driven by $z_\pm \neq 0$ with $x_\pm = 0$, forming displaced coherent states along $X_\alpha$. These attractors correspond to clockwise flows in the closed system (minima of the Ginzburg-Landau potential), which turn into spirals under dissipation.

For $U > 0$ (hardening nonlinearity), the uncoupled cavity supports a stable high-amplitude regime, and similar superradiant physics. In contrast, for $U < 0$, cavity resonance shifts to lower frequencies, making the system prone to instability as the restoring force weakens. Without coupling, the cavity could become trapped in a self-amplifying cycle, leading to runaway growth (see Fig. 1(b)). However, backaction in this regime further softens the restoring force, stabilizing the system and giving rise to the KP phase.

\renewcommand{\refname}{Supplemental References}
\hypersetup{urlcolor=blue}

\hypersetup{urlcolor=blue}

\end{document}